\documentclass[11pt]{article}
\usepackage{moriond,epsfig}

\def\be{\begin{equation}}
\def\ee{\end{equation}}
\def\bea{\begin{eqnarray}}
\def\eea{\end{eqnarray}}

\begin{document}
\vspace*{4cm}
\title{LHC: status and commissioning plans}

\author{ M. Lamont }

\address{CERN, Geneva, Switzerland}

\maketitle\abstracts{In 2008 the LHC saw a series of injection tests with beam and the start of full beam commissioning. Initial beam commissioning went well and circulating beam was quickly established. Progress was unfortunately curtailed by the sector 34 incident which caused extensive damage.
The causes of the incident are recalled and the status of repairs and consolidation  measures are presented. The schedule and luminosity targets for the 2009 - 2010 run are discussed. }

\section{Introduction}

After an intense period of preparation, the LHC started full beam commissioning on September 10th 2008. Initial progress was excellent, however commissioning was curtailed by the serious incident in sector 34 which saw the opening of an interconnect busbar during powering tests without beam. This provoked a destructive release of a large volume of Helium into the insulation vacuum of the sector. The damage caused was extensive and has required the removal of 51 main magnets (dipoles and quadrupoles) and the repair of the considerable collateral damage.

This paper gives a brief overview of the present understanding of the causes of the incident; the status of the repair; the compensatory measures that are being put in place to ensure that the incident never recurs and those that aim to alleviate the effects of any future problems. It is hoped to start with beam again at the end of September 2009 and run through the winter to provide the experiments with a significant integrated luminosity at a target beam energy of 5 TeV. 

\section{Sector 34}

\subsection{Cause of sector 34 Incident}\label{subsec:prod}

The current of the main magnet circuits passes through busbars in the interconnects between quadrupole and dipole and between dipole and dipole. These busbars consist of the superconducting Rutherford cable soldered inside a rectangular cross-sectioned copper envelope~\cite{av}. 

The busbars from adjacent magnets are spliced together by  soldering overlaying protruding lengths of Rutherford cable inside segments of copper to produce a continuous busbar with good connectivity between neighbouring segments of copper and a well soldered Rutherford cable joint. 

The present understanding is that a splice with bad thermal and electrical contact between the superconductor and the copper produced sufficient resistive heating to lead to thermal runaway. This provoked the melting of the material surrounding the splice, and subsequently an electric arc developed between the two exposed cable ends. This arc melted through the helium line in which the cable travels, releasing Helium into the insulation vacuum of the interconnect. The rapid and voluminous expansion of the Helium caused a pressure wave that propagated along the insulation vacuum causing extensive damage~\cite{plb}.

Subsequent examination of cryogenics temperature measurements led to the conclusion that the splice had a resistance of 220 nano-ohm compared with the nominal value of around 0.3 nano-ohm.

\subsection{Sector 34 repair}

Damage to the sector was extensive and has necessitated a full-scale repair program, briefly:

\begin{itemize}\addtolength{\itemsep}{-0.5\baselineskip}

    \item Removal of the damaged magnets: 53 main magnets in total, 39 dipoles, 14 quadrupoles and their associated correctors (these are incorporated into a short straight section or SSS).
\item Repair of damage to the cryogenics supply line.
\item Cleaning of  beam vacuum which was polluted locally with soot and throughout the sector with debris from the magnets' super-insulation.
\item Repair of damaged beam vacuum segments.
\item Refurbishment of lightly damaged magnets and the configuration and preparation of spare magnets.
\item Cool-down and tests of all magnets on the surface before re-installation.
\item Re-installation of magnets.
\item Re-interconnection and a  program to address all nonconformities.
\item Execution of rigorous electrical and assembly quality assurance.
\item Cool-down of the sector with associated electrical quality assurance.
\item Re-testing of all circuits.

\end{itemize}
 
At the time of writing (end May 2009), all magnets have been re-installed, most of the interconnect work has finished and the thorough program of quality assurance tests are in progress.

The sector 34 repair represents a huge effort by a number of different teams.
Quality assurance has been pursued rigorously and a number of issues, problems and nonconformities have been addressed.

\section{Consolidation and Compensatory Measures}

\subsection{Extended Quench Protection System}

An additional layer of detection electronics will ensure
the safe detection of aperture symmetric quenches by
comparing the total voltage drops across magnets. The
system will share the instrumentation cables with the
suspicious splice detection system and evaluate the
signals derived from four dipole magnets and two
quadrupole magnets. 

The Quench Protection System (QPS)
will also be extended  to detect and localize
suspect splices which could represent a potential risk for
the LHC main circuits (RB, RQD
and RQF)~\cite{rd}. According to the outcome of recent
simulations \cite{av} a splice developing a resistance in the
order of 50 to 100 nano-ohm is regarded as potentially dangerous
at higher currents. This value is equivalent to a voltage
drop of 100 mV or a generated power of 0.1 W at a
current of 1 kA. In such a case the system should
interlock the concerned circuits by initiating a fast
discharge by activation of the energy extraction systems.

The necessary detection threshold has been determined to
300 micro-volt with 10 s evaluation time. In addition the
system will provide data for enhanced diagnostics via the
QPS supervision allowing a measurement of the splice
resistance with a resolution of about 1 nano-ohm.
As the new system can only access the superconducting
circuits via the existing voltage taps routed to the magnet
interface box connectors (diode voltage taps) it is not possible to provide coverage of a single splice.

The extension of the QPS system represents a major upgrade and necessitates the design, prototyping, production of a large number of new boards and crates. Full scale cabling is required in the LHC tunnel and a full series of component and systems tests will be required. The installation and commissioning of the system is very much on the critical path for LHC re-commissioning in 2009.

\subsection{Helium release - compensatory measures } 

In order to minimize the collateral damage should a similar incident to that in sector 34 occur, a number of measures have been taken that anticipate the so-called Maximum Credible Incident (MCI)~\cite{js}. The MCI foresees the rupture of all three Helium lines in a interconnect and a peak Helium discharge of 40 kg/s. Measures to deal with a potential MCI include:
\begin{itemize}\addtolength{\itemsep}{-0.5\baselineskip}
    \item New relief valves (DN200) - these have been installed on each dipole of four LHC sectors in 2009. The intervention has to be done at warm and these sectors were brought up to room temperature at the end of 2008.

\item The sectors that have remained cold in 2009 have had their instrument ports in the SSS equipped with springs, providing an additional Helium release pathway - although not enough to deal with the full MCI. These four remaining sectors will be equipped with DN200 relief valves in the 2010 shutdown.

\item Relief valves have also been fitted on stand alone magnets in the long straight sections and on the various current feed boxes.

\item New support jacks have been installed on the SSS with vacuum barriers. These jacks are able to withstand the forces generated in the case of a MCI preventing lateral movement and  knock-on damage~\cite{oc}.

\end{itemize}

\subsection{Other consolidation }

Numerous other fixes and consolidation measures have taken place in the forced extended shutdown. This work has included resolution of problems with the so-called continuous cryostats (empty cryostats in the dispersion suppressors); movement of electronics susceptible to radiation; decrease of dipole and quadrupole energy extraction time; re-configuration the UPS system etc.

\section{Splices}

\subsection{Resistive Spices}

A number of techniques~\cite{ncl} have been developed to detect splices with potentially dangerous resistance, with the stated aim to measure all splices cold. (At low temperature the current flows through the superconductor and one is attempting to measure the resistance of the soldered joint between the Rutherford cable segments.) The methods are listed below.

\begin{itemize} \addtolength{\itemsep}{-0.5\baselineskip}
    \item The Keithley method which uses a nanovoltmeter on specific segments previously identified as possibly problematic.
 
\item The QPS snapshot method which uses QPS diagnostic data to measure the voltage drop across a low number of splices. Using this technique around 6600 splices have be measure and two bad splices found.
\item Calorimetry - cryogenics temperature monitoring measurements in cold sectors have revealed potential problems in splices inside magnets. 57\% of all splice have been covered by this method with two bad splices found and confirmed.
\item Ultrasound and gamma source techniques have been principally used in quality control and splice diagnostics. Gamma analysis has revealed the presence of "voids" where solder has been drained from the magnet side busbar during the splice brazing process.
\item Measurement data taken during the surface tests of all magnets has also been reanalyzed.
\end{itemize}

Splice resistance nonconformities to date include a 200 nano-ohm interconnect case (the cause of the sector 34 incident), a magnet inter-pole splice of 100 nano-ohm; a magnet inter-pole splice of  50 nano-ohm; a magnet inter-aperture splice of 30 nano-ohm. About half the machine has been checked cold to 40-60 nano-ohm and third to better than 20 nano-ohm.  The other half of the machine and still to be measured (and the first half re-measured with better accuracy).

\subsection{Warm splice measurements}

One additional danger that has recently surfaced is a bad electrical contact between the copper of the busbar and the U-profile of the splice insert on at least one side of the joint. Combined with a bad contact between the cable and the copper this leaves the splice without an alternate route for the current in the case of a busbar quench - in a good splice the current can flow in the copper removing the danger of excessive resistive heating in the quenched superconductor. A good contact between the Rutherford cable joint is assumed (i.e. less that 2 nano-ohm)~\cite{bf}.

Such situation can be detected by measurements at warm using low current and a nanovoltmeter across short segments of the machine. Under such circumstances the current flows in the copper and the resistance of a good joint is around 12 micro-ohm. Extensive measurements of the four warm sectors (May 2009) have revealed 16 segments with excess resistance of over 30 micro-ohm. The relevant interconnects have been opened. Individual splice measurements have revealed resistances of 30 - 50 micro-ohms. All such splices have been re-done and re-measured.

Warm quadrupole measurements started in May 2009. Measurements at 
at 80 K in sector 23 are also ongoing at this time. The measurements at 80 K are more difficult and show a lot more signal variation - the resistivity of copper falls by a factor of 7.5 at this temperature. The question of what to do if suspect splices are found at this stage of re-commissioning is to be addressed.

\section{2009 - 2010 run}

\noindent The baseline for the 2009 - 2010 run was established at Chamonix and foresees
one month commissioning to establish first collisions, a ten month proton physics run followed by a one month Lead ion run with shutdown foreseen at the end of September 2010. The run will be through the winter with a possible technical stop under discussion over Christmas~\cite{sm}.

The curtailed but productive beam commissioning period in 2008 coupled with the results from the injection tests have given us some confidence in numerous aspects of the LHC's potential operation, namely: magnet model, magnet field quality, machine aperture, machine alignment, optics, injection and beam dump systems, collimation, beam instrumentation, controls and software~\cite{ml}.
It also had to be noted that only limited progress was made into the full beam commissioning program and that a lot remains to be done before we reach the first major milestone of the program - colliding low intensity beams at high energy. Systematic and careful progress will be essential if the potential dangers of even moderate intensity beams are to be dealt with properly. It is estimated that approximately four weeks will be required  to establish first collisions given reasonable machine availability. These collisions will be with un-squeezed, low intensity beams.

Commissioning will continue thereafter for some weeks as the intensity is increased, the squeeze commissioned and the machine protection and other system commissioning is performed. This stage can be interleaved with pilot physics with a gradual increase in intensity, a lowering of $\beta^{*}$ and a concomitant increase in the luminosity. Luminosity will remain modest during this phase.

\subsection{Luminosity}

The stated aim of the 2009-2010 run~\cite{sm} is to deliver a few hundred~pb$^{-1}$ of good data to the general purpose detectors. 
Issues that should be considered in estimates of possible integrated luminosity are listed below.
\begin{itemize}\addtolength{\itemsep}{-0.5\baselineskip}

 \item The total beam intensity achievable during this period will depend critically on the  cleaning efficiency of collimation system and the to be established quench limits. Recent estimates ~\cite{ralph} put the total current limits at  2e13 protons for intermediate collimator settings and, after some  experience, 5e13 for tight collimator settings. Such limitations clearly impact the maximum achievable luminosity.

\item At 5 TeV there is damage potential with even low beam intensities. The machine protection must be fully tested and qualified before any move to higher intensities.

 \item The number of bunches will be increased from 
2 bunches on 2  bunches (2x2), and increase through 43x43 (43-4-43-19), to 156x156 (156-4-16-72). The second set of figures representing the number of colliding bunches in Atlas, Alice, CMS and LHCb respectively. Up to 156 bunches there are no parasitic encounters either side of the experiments and the crossing angles can remain off.

\item The bunch intensity can be increased from 5e9 to 11e10. Given a limit on the total intensity it is of interest to use the minimum number of bunches.

\item Above 156 bunches per beam the  crossing angle must be brought on. This is not necessarily cost free and consideration of related effects such as beam-beam, dynamic aperture, physical aperture must be made.

\item Running below 7 TeV brings bigger beams and less physical aperture.

\item $\beta^{*}$: The limit at 5 TeV is  3  to 1 m in Atlas and CMS~\cite{wh}. Bringing the crossing angle on pushes the minimum $\beta^{*}$ up. 
LHCb has the additional restraint of their spectrometer bump and consequently are limited to 4 to 2 m.

\item Many machine operations issues such as 
 injection, ramp, squeeze, beam lifetime,  background optimization in physics have to be mastered.

\item Machine availability will be critical; the problem space of the LHC is huge and low availability is a real possibility in the first months. Time must also be foreseen for 
continued beam commissioning, machine development, access etc.

\end{itemize}

It's clear that possibilities exist to optimize luminosity parameter space given the above limitations and experience.   $1 \times 10^{32}$~cm$^{-2}$~s$^{-1}$ would be an encouraging achievement and integrated luminosity of 200-300~pb$^{-1}$ of good data would appear a reasonable target~\cite{ml2}.

\section{Conclusions}

The seriousness of the sector 34 incident can not be understated. There has been a remarkable effort to effect a repair. Considerable measures have been taken to mitigate the effects of any similar incident and to make sure something like it never happens again. 

There are thousands of splices in the LHC and some still have be measured either warm or cold. A concerted program of measurements is ongoing. There is now good understanding of the causes of the sector 34 incident and of other possibilities of failure.

First experience with beam in the LHC was encouraging. The 2009 - 2010 run foresees first beam around the end of September 2009, first collisions at high energy with a month, followed by 10 months of proton physics. The aim is to deliver a few hundred~pb$^{-1}$ of good data to the general purpose detectors. 

\section*{Acknowledgments}
The LHC is huge enterprise of unparalleled complexity. The work following the sector 34 has been, and continues to be, intense and represents a committed effort from everyone at CERN and many collaborations around the world.

\end{document}